\documentclass[9pt,twocolumn,twoside]{pnas-new}

\templatetype{pnasresearcharticle} 

\usepackage{amssymb,amsfonts,amsmath,amsthm}
\usepackage{graphicx}
\usepackage{bm}
\usepackage{mathrsfs}
\usepackage{esint}
\usepackage{enumitem}
\usepackage{multirow}
\usepackage[normalem]{ulem}
\usepackage{xcolor}
\usepackage{hyperref}
\hypersetup{
    colorlinks=true,                          
    linkcolor=red, 
    citecolor=red, 
    urlcolor=red  }
\usepackage{placeins} 
\renewcommand\b[1]{{\bf  #1}}
\renewcommand\vec[1]{\boldsymbol{#1}}

\renewcommand\rm{\mathrm}
\newcommand\tr{\mathrm{tr}}
\newcommand\del{\nabla}

\newcommand\dd{\mathrm{d}}

\renewcommand\ss[1]{{\color{black} #1}}

\title{Design rules for controlling active topological defects}

\author[a,b,1,2]{Suraj Shankar}
\author[c,d,1]{Luca V.~D.~Scharrer}
\author[c,e]{Mark J.~Bowick}
\author[c,2]{M.~Cristina Marchetti}

\affil[a]{Department of Physics, Harvard University, Cambridge, MA 02138, USA}
\affil[b]{Department of Physics, University of Michigan, Ann Arbor, MI 48109, USA}
\affil[c]{Department of Physics, University of California Santa Barbara, Santa Barbara, CA 93106, USA}
\affil[d]{Department of Physics, The University of Chicago, Chicago, IL 60637, USA}
\affil[e]{Kavli Institute for Theoretical Physics and Department of Physics, Santa Barbara, CA 93106, USA}

\leadauthor{Shankar} 

\significancestatement{\ss{Active fluids, such as bacterial suspensions and flowing tissues, often exhibit orientational order disrupted by topological defects. Much is understood about how internal driving powers nonequilibrium dynamics, but how can we design protocols to transport and organize patterns in active fluids for functional goals? We develop an additive symmetry-based framework to control the dynamics of such topological defects by spatio-temporally manipulating active stresses. Our framework identifies design principles for controlling defect trajectories using active tweezers, as well as collections of interacting defects using patterns of activity. By combining simulations with theory, we uncover necessary symmetry conditions and trade-offs that govern defect control policies in active media, suggesting general rules for manipulating a broad range of synthetic and biological active matter.
}}

\authorcontributions{}
\authordeclaration{The authors declare no competing interests.}
\equalauthors{\textsuperscript{1}S.~S.~and L.~V.~D.~S~contributed equally to this work.}
\correspondingauthor{\textsuperscript{2}Correspondence and requests for materials should be addressed to M.~C.~M.~Email: cmarchetti@ucsb.edu~, S.~S.~Email: surajsh@umich.edu~.}


\begin{abstract}
Topological defects play a central role in the physics of many materials, including magnets, superconductors and liquid crystals.
In active fluids, defects become autonomous particles that spontaneously propel from internal active stresses and drive chaotic flows stirring the fluid. The intimate connection between defect textures and active flow suggests that properties of active materials can be engineered by controlling defects, but design principles for their spatiotemporal control remain elusive. Here we {propose} a symmetry-based additive strategy for using elementary activity patterns, as active topological tweezers, to create, move and braid such defects. By combining theory and simulations, we demonstrate how, at the collective level, spatial activity gradients act like electric fields which, when strong enough, induce an inverted topological polarization of defects, akin to {a negative} susceptibility dielectric. We harness this feature in a dynamic setting to collectively pattern and transport interacting active defects. Our work establishes an additive framework to sculpt flows and manipulate active defects in both space and time, paving the way to design programmable active and living materials for transport, memory and logic.
\end{abstract}

\dates{This manuscript was compiled on \today}
\doi{\url{www.pnas.org/cgi/doi/10.1073/pnas.XXXXXXXXXX}}

\begin{document}

\maketitle
\thispagestyle{firststyle}
\ifthenelse{\boolean{shortarticle}}{\ifthenelse{\boolean{singlecolumn}}{\abscontentformatted}{\abscontent}}{}

\dropcap{T}he ability to manipulate matter at the nano, micro and mesoscale is essential for developing functional materials with adaptive and responsive properties \cite{xia2022responsive}. A common strategy is to employ repetitive assembly of discrete physical units (e.g., polymers, colloids, elastic beams etc.) to construct (meta)materials that acquire novel morphologies and functionalities from their architecture \cite{jacobs2016self,dey2021dna,zeravcic2017colloquium,bertoldi2017flexible,xia2022responsive}. When rationally designed, such structured materials exhibit unconventional mechanical responses \cite{bertoldi2017flexible}, enable programmable computation \emph{in materia} \cite{zangeneh2021analogue,yasuda2021mechanical} and mimic living systems~\cite{zeravcic2017colloquium}. {But this approach doesn't apply to systems with mesoscale order that behave as continuous media. An alternate strategy is to use naturally localized excitations, such as topological defects, as discrete building blocks of a hierarchical material.}


Topological defects are characteristic singularities that emerge when ordered phases of matter are {rapidly quenched} or are frustrated by boundaries and external fields \cite{chaikin2000principles}. {These} discrete excitations {encode} information of the continuous order parameter in robustly localized singularities that behave as effective quasiparticles. From manipulating magnetic skyrmions \cite{romming2013writing,fert2017magnetic} 
and superconducting vortices \cite{berciu2005manipulating,ge2016nanoscale} to knotting disclination loops in liquid crystals \cite{tkalec2011reconfigurable,tai2019three,foster2019two},
the control of defects in diverse systems offers new opportunities for designing reconfigurable memories and logic devices.

\begin{figure*}[t]
    \centering{\includegraphics[width=0.75\textwidth]{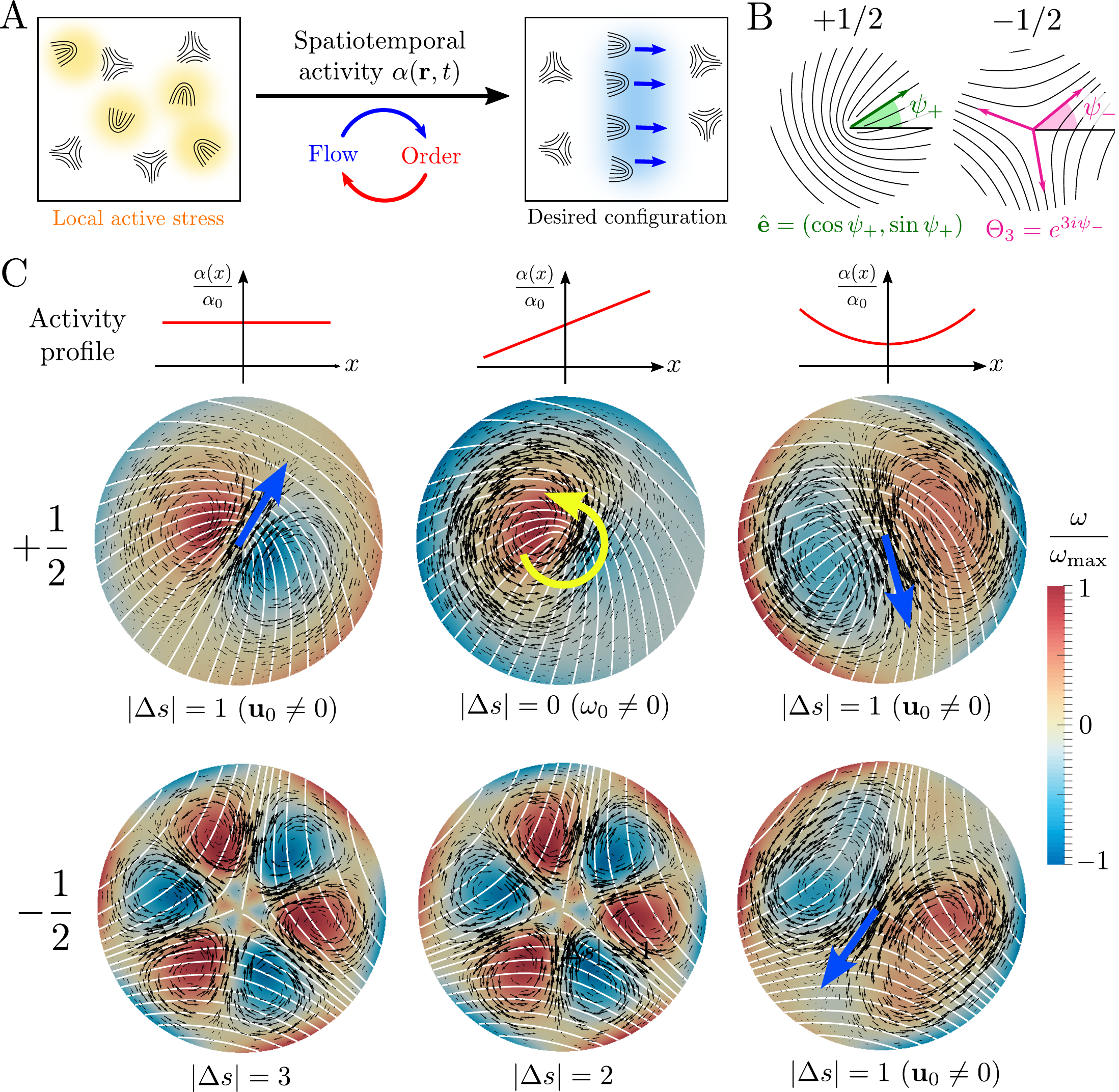}}
    \caption{{\bf Additive framework for spatiotemporal control of active defects.} (A) Active stresses generate flow through distortions of order in active nematic fluids resulting in the proliferation of motile defects. How can we locally actuate active stresses to achieve desired defect textures and flow patterns in space-time? (B) Disclinations with topological charge $\nu=\pm 1/2$ have distinct symmetries characterized by a vector $\hat{\b{e}}$ ($+1/2$) and a complex triatic parameter $\Theta_3$ ($-1/2$). (C) In the vicinity of a defect (director, white lines), simple polynomial activity profiles (constant: left, linear: middle, quadratic: right) in space locally generate distinct flow ($\b{u}$, black arrows) and vorticity ($\omega=\hat{\b{z}}\cdot(\vec{\del}\times\b{u})$, heat map) fields, here computed using Eq.~\ref{u} with a screening length $\ell_\eta=\sqrt{\eta/\Gamma}=2\xi$ and no-slip boundary conditions. The nature of the flow velocity and vorticity at the defect core ($\b{u}_0$, blue arrow; $\omega_0$, yellow arrow) is dictated by the combined rotational symmetry of the defect texture ($s$) and the activity pattern ($n$) quantified by the absolute index difference $|\Delta s|=|s-n|$. Linear combinations of individual activity patterns simply sum the respective flow fields providing an additive strategy for generating arbitrarily complex active flows.}
    \label{fig1}
\end{figure*}

Topological defects also naturally emerge in \emph{active} fluids \cite{marchetti2013hydrodynamics,doostmohammadi2018active,shankar2022topological}, i.e., fluids composed of self-driven units such as bacteria, motor protein-biofilament constructs, active colloids and living cells \cite{zhou2014living,li2019data,sanchez2012spontaneous,bricard2013emergence,duclos2018spontaneous} that can organize into states with (nematic) liquid crystalline order. {Elementary defects in two-dimensional (2D) nematics are disclinations characterized by a topological winding number $\nu=\pm 1/2$ \cite{chaikin2000principles}. In active nematics, spontaneous flows are driven by self-propelled $+1/2$} defects \cite{giomi2013defect,shankar2018defect,shankar2022topological} that chaotically self-stir the fluid \cite{tan2019topological,alert2022active,serra2023defect}. The intimate feedback between distortions of order, dynamical defects and flow (Fig.~\ref{fig1}A) makes active fluids an attractive platform for manipulating transport of matter, energy and information far from equilibrium \cite{needleman2017active,zhang2021autonomous,shankar2022topological}. Global control of active flows have been achieved by imposing constraints via geometry, confinement, substrates, etc.~\cite{peng2016command,opathalage2019self,keber2014topology,ellis2018curvature,guillamat2016control,thijssen2021submersed}. More recent experimental advances in optically responsive platforms now allow \emph{local} spatiotemporal control of internal stresses in bacterial and synthetic active fluids \cite{dervaux2017light,arlt2018painting,frangipane2018dynamic,ross2019controlling,zhang2021spatiotemporal,linnea2022} enabling unprecedented abilities to sculpt and configure active materials on demand. Complementing these experimental results, recent theoretical works have also begun exploring the inverse problem in the context of optimal control \cite{norton2020optimal,shankar2022optimal}, reinforcement learning \cite{falk2021learning} and pattern formation \cite{shankar2019hydrodynamics,mozaffari2021defect,ruske2022activity,yang2022}.

{But how can we construct spatiotemporal profiles of active stresses to dynamically control active defects and flow (Fig.~\ref{fig1}A)?}
{Current approaches largely focus on controlling the motion of individual $+1/2$ defects by using geometric patches of constant activity \cite{zhang2021spatiotemporal,zhang2022logic,mozaffari2021defect}. Optimal control policies generate complex dynamical activity profiles \cite{norton2020optimal} that often lack clear interpretation. Other studies have explored the response of $+1/2$ defects to linear ramps and step jumps in activity \cite{tang2021alignment,jonas2022,mozaffari2021defect,zhang2021spatiotemporal}, but a systematic quantification of defect patterns as a function of activity gradients is absent (see Refs.~\cite{shankar2019hydrodynamics,ruske2022activity} for recent efforts). Furthermore, none of the existing approaches allow $-1/2$ defects to be controlled. A rational design framework to control both $\pm1/2$ topological defects using activity is lacking.}


\ss{Here we systematically solve this inverse problem by developing a symmetry-based additive approach (Fig.~\ref{fig1}) that exploits the linearity of inertia-less dynamics. A key result is the derivation of a selection rule which provides the necessary symmetry conditions that an activity pattern must satisfy to generate any desired $\pm 1/2$ motion. The usefulness of these rules is demonstrated in numerical simulations by designing active topological tweezers that can manipulate complex defect trajectories. At the many-body level, a hydrodynamic approach generalizes these results to include defect interactions and helps understand both the static and dynamic response of an interacting defect gas to activity gradients, that we numerically quantify. Altogether, our framework unifies existing results and provides general design principles for controlling arbitrary charge defects in uniaxial active fluids.}


\section*{Nematodynamics with spatiotemporal activity}
We model the active fluid as a thin, two dimensional (2D) viscous layer with a nematic order parameter $\b{Q}$ and flow velocity $\b{u}$ whose coupled dynamics is governed by Stokesian force balance and passive relaxation, which give \cite{marchetti2013hydrodynamics,doostmohammadi2018active,shankar2022topological}
\begin{align}
    &\partial_t\b{Q}+\b{u}\cdot\vec{\del}\b{Q}=\b{S}(\b{u},\b{Q})+\dfrac{1}{\gamma}\b{H}\;,\label{Q}\\
    &-\Gamma\b{u}+\eta\del^2\b{u}-\vec{\del}\Pi+\vec{\del}\cdot\left(\vec{\sigma}^a+\vec{\sigma}^{el}\right)=\b{0}\;.
\label{u}
\end{align}
Local order is advected, rotated and sheared by flow (captured by $\b{S}(\b{u},\b{Q})$, see Methods for details) and distortions relax via the molecular field $\b{H}=(a_2-a_4\tr[\b{Q}^2])\b{Q}+K\del^2\b{Q}$ ($a_{2,4}>0$) with elasticity $K$ and rotational viscosity $\gamma$ (Eq.~\ref{Q}). Force balance (Eq.~\ref{u}) includes damping from viscosity ($\eta$) and friction ($\Gamma$), pressure $\Pi$ enforcing incompressibility ($\vec{\del}\cdot\b{u}=0$), an elastic stress $\vec{\sigma}^{el}$ (see Methods for details) and an active stress $\vec{\sigma}^a=\alpha\b{Q}$ \cite{marchetti2013hydrodynamics}, where the activity $\alpha$ captures the average strength of the oriented force dipoles exerted by the active units on the fluid ($\alpha<0$: extensile, $\alpha>0$: contractile). Importantly, activity $\alpha(\b{r},t)$ varies in space and time, serving as the control variable and we focus on the extensile case ($\alpha<0$) relevant for most experimental systems \cite{sanchez2012spontaneous,doostmohammadi2018active} (though our results hold more generally). In the following we work in units such that the nematic correlation length $\xi=\sqrt{K/a_2}=1$, the nematic relaxation time $\tau_n=\gamma/a_2=1$, and a passive elastic stress scale $K\Gamma/\gamma=1$


\section*{Controlling isolated active defects}
\subsection*{Symmetry-based selection rule}
As any distortion of order generates flow in an active fluid (Eq.~\ref{u}), defects self-generate local active flows that advect and rotate them \cite{giomi2013defect,pismen2013dynamics,shankar2018defect}. By assuming a separation of timescales wherein nematic distortions relax faster than the dynamics of defects \cite{giomi2013defect,pismen2013dynamics,shankar2018defect}, we can neglect the nonlinearity in Eq.~\ref{Q}. Thus to control the trajectory of an individual defect described by $3$ independent degrees of freedom (DoFs, two translational, one rotational), we need to prescribe the local flow velocity $\b{u}_0$ and vorticity $\omega_0$ actively generated at the defect core. But the control parameter, the activity $\alpha(\b{r},t)$, is a single scalar \emph{field} with effectively infinite DoFs. To overcome this DoF mismatch, reminiscent of similar problems in neuromotor control \cite{bernshtein1967co}, we turn to symmetry.

{Both $\pm1/2$ defects are distinguished by the geometry of their local nematic texture; comet-shaped $+1/2$ defects have a polarity captured by a unit vector $\hat{\b{e}}$ (Fig.~\ref{fig1}B, left), while triangular $-1/2$ defects have a three-fold symmetry captured by a unit complex triatic parameter $\Theta_3$ (Fig.~\ref{fig1}B, right) \cite{vromans2016orientational,tang2017orientation}; see SI for details.}
In the simple case of constant activity, it is well-known that $+1/2$ defects self-propel along their polarity with nonzero flow at their core ($\b{u}_0^+\propto \alpha\hat{\b{e}}$) while three-fold symmetric $-1/2$ defects do not ($\b{u}_0^-=\b{0}$) \cite{doostmohammadi2018active,shankar2022topological}, highlighting the importance of defect rotational symmetry for its motion. For a generic activity profile, because the active stress $\vec{\sigma}^a$ is bilinear in activity ($\alpha$, the control) and the nematic texture ($\b{Q}$, the state), the combined symmetry of the two fields dictates the nature of local flow. Euclidean isometries of the plane that leave the origin (defect core) fixed are characterized by the orthogonal group $\mathsf{O}(2)$ and all its discrete dihedral subgroups $\mathsf{D}_n$ (i.e., the symmetries of an $n$-sided regular polygon). We thus expand the activity in terms of angular Fourier harmonics ($\alpha(\b{r})=\sum_{n}\tilde{\alpha}_n(r)e^{in\phi}$, $\phi$ is the polar angle) which offer a natural basis with definite $n$-fold dihedral symmetry. 
The rotational symmetry of 2D nematic defects can be similarly quantified. Defect textures with topological winding $\nu$ can be assigned an integer symmetry index $s=2|1-\nu|$ \cite{tang2017orientation} corresponding to their dihedral symmetry group $\mathsf{D}_s$. As expected, $+1/2$ defects have $s=1$ and $-1/2$ defects have $s=3$.


\begin{figure*}
    \centering{\includegraphics[width=0.8\textwidth]{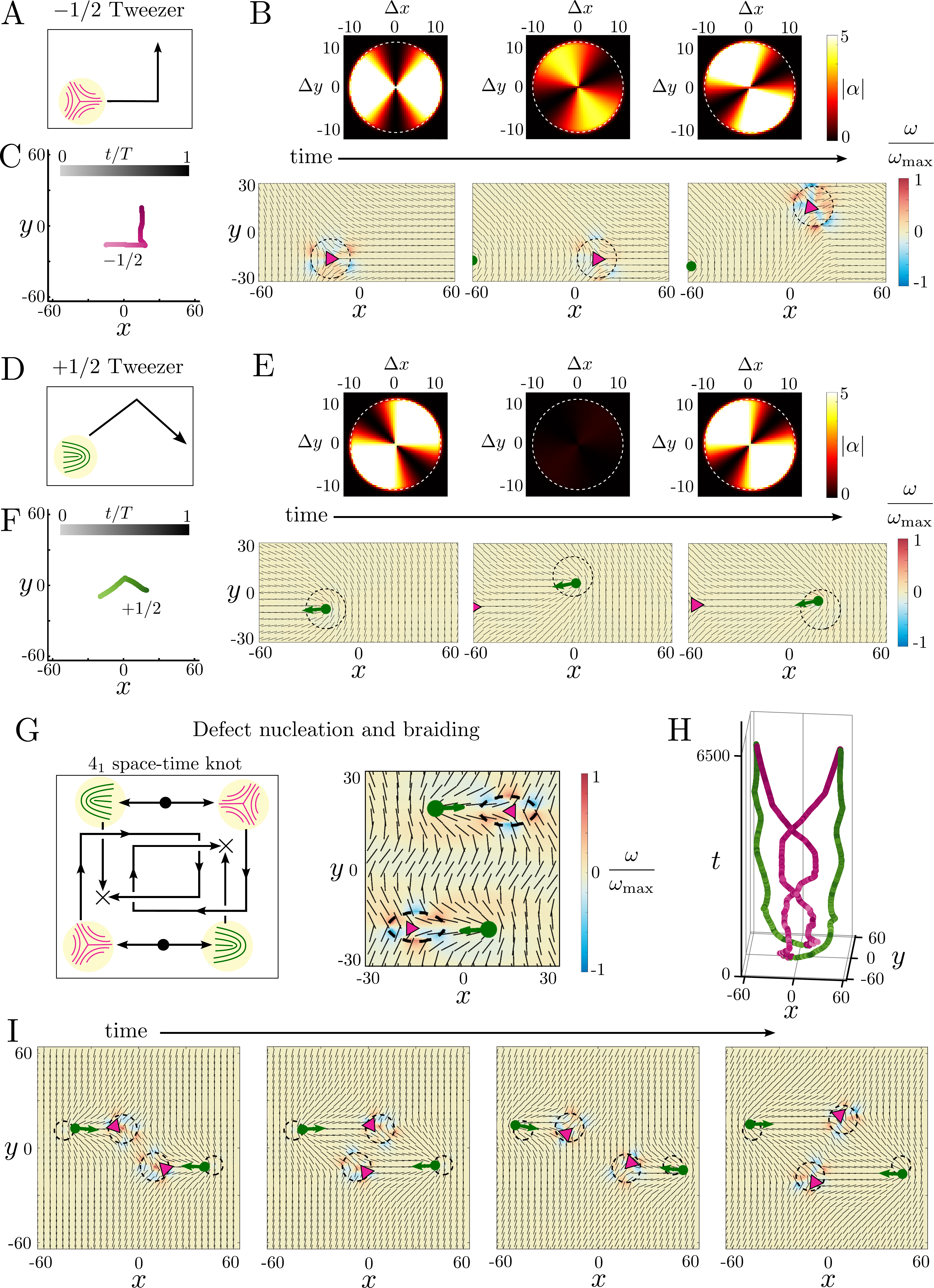}}
    \caption{{\bf Active topological tweezers enable complex manipulation of defect paths.} (A-C) Moving a $-1/2$ defect using an active tweezer; see Movie S1 (path shown in A). (B) The $-1/2$ defect (magenta triangle) tracks the motion of a small activation disc (dashed circle) with a centered quadrupolar activity profile (see Eq.~\ref{eq:tweezer} and SI for details) whose orientation controls the local flow and direction of motion. $(\Delta x,\Delta y)$ denote distance from the core of the controlled defect. (C) Tracked trajectory of $-1/2$ defect shaded by the normalized time $t/T$ ($T$ is the protocol duration). (D-F) Moving a $+1/2$ defect using an active tweezer; see Movie S2 (path shown in D). (E) The $+1/2$ defect (green arrow) tracks the motion of a small activation disc (dashed circle) with a centered quadrupolar activity profile (see Eq.~\ref{eq:tweezer} and SI for details). (C) Tracked trajectory of $+1/2$ defect shaded by normalized time $t/T$. (G) Demonstration of simultaneous multidefect control using active tweezers, with defect pair creation (filled circles, left), braided trajectories and pair exchange (arrows, left), ending in pair annihilation (crosses, left); see Movie S3. An elliptic localized patch of high activity allows local nucleation of $\pm1/2$ defect pairs (right, zoomed-in snapshot). (H) Tracked trajectory of defects ($+1/2$: green, $-1/2$: magenta) forms a closed braid, the figure-eight ($4_1$) knot, in space-time. (I) Snapshots showing defect braiding in time.}
    \label{fig2}
\end{figure*}

To obtain a finite defect velocity or vorticity ($\b{u}_0,\omega_0\neq 0$), the angle average (zeroth angular moment) of the respective fields must be nonvanishing.
Linearity of Eq.~\ref{u} and the form of the active stress $\vec{\sigma}^a$ then prescribe a simple selection rule (see Theorem~3.1 in SI for proof): for a defect with symmetry $s$ subjected to an $n$-fold symmetric activity profile (only $\tilde{\alpha}_{\pm n}\neq 0$), a necessary condition for self-propulsion ($\b{u}_0\neq 0$) is $|s-n|=1$, and for self-rotation ($\omega_0\neq 0$) is $|s-n|=0$ (see Fig.~\ref{fig1}C).
While symmetry dictates the existence (or not) of local active flow, its specific form and direction depend on details of the activity pattern and defect orientation.  
An explicit calculation of $\b{u}_0$, $\omega_0$ for $\pm 1/2$ defects in a smoothly varying activity profile yields (see SI for details)
\begin{align}
    &\b{u}_0^+=\vec{\mathcal{V}}_+\cdot\hat{\b{e}}\;,\quad\omega_0^+=\hat{\b{z}}\cdot\left(\vec{\Omega}_+\vec{\times}\hat{\b{e}}\right)\;,\label{uw1}\\
    &\b{u}_0^-=\left(\rm{Re}[\Theta_3\mathcal{V}_-],\,\rm{Im}[\Theta_3\mathcal{V}_-]\right)\;,\quad\omega_0^-=-\rm{Re}[i\Theta_3\Omega_-]\;,\label{uw2}
\end{align}
where (to leading order in gradients) $\vec{\mathcal{V}}_+\propto\alpha\b{I}+\mathcal{O}(\del^2\alpha)$, $\mathcal{V}_-\propto\partial^2\alpha$ ($\partial=(\partial_x-i\partial_y)/2$), $\vec{\Omega}_+\propto\vec{\del}\alpha$ and $\Omega_-\propto\partial^3\alpha$ are translational and rotational response coefficients that depend linearly on the local activity and its symmetry mandated gradients evaluated at the defect core (see SI for details). In Eqs.~\ref{uw1},~\ref{uw2}, we have simply Taylor-expanded the activity near the defect core upon assuming weak spatial gradients. Flows generated by $\pm1/2$ defects in elementary polynomial profiles of activity are shown in Fig.~\ref{fig1}C. 
Strikingly, we see that while linear gradients of activity ($\vec{\del}\alpha\neq\b{0}$; $n=1$ symmetry) only contribute to vorticity of $+1/2$ defects, consistent with \cite{tang2021alignment,ruske2022activity,jonas2022}, quadratic activity gradients ($\vec{\del}\vec{\del}\alpha\neq\b{0}$; $n=2$ symmetry) generate net active flow for both $\pm1/2$ defects (Fig.~\ref{fig1}C) as predicted by the symmetry selection rule. Arbitrary linear combinations of activity patterns then simply superpose their respective flow fields thereby providing an attractive additive and modular strategy for designing complex local flows from active defects.

\subsection*{Active topological tweezers}
We now deploy our design framework in numerical simulations of Eqs.~\ref{Q},~\ref{u} (see Methods and SI for details) to construct activity patterns for basic defect based operations. For controlling individual defects, we generalize ``topological tweezers'' \cite{irvine2013dislocation} used in colloidal crystals to devise \emph{active topological tweezers} (Fig.~\ref{fig2}), i.e., a disc of finite activity, whose motion and local spatiotemporal pattern is chosen to achieve a prescribed defect transport task.

\ss{
The tweezer design is constrained by the following three conditions - (i) the selection rule for the symmetry of the activity pattern, (ii) that the maximum activity is smaller than the bend-instability threshold ($|\alpha|\lesssim (K\Gamma/\gamma)\min(1,(\pi\ell_\eta/R)^2)$, where $\ell_\eta=\sqrt{\eta/\Gamma}$ is the screening length and $R$ is the tweezer radius) and (iii) $\alpha$ does not change sign (so the system never switches from extensile to contractile or vice-versa). To reduce the effort of actuation, we minimize gradients in activity, so for simplicity, we choose radially constant activity profiles as far as we can and choose the smallest angular variation required to satisfy the selection rule. These design principles allow the construction of simple topological tweezers for translating $\pm1/2$ defects.

The selection rule dictates that a quadrupolar or $2$-fold symmetric activity profile ($\tilde{\alpha}_{\pm2}\neq 0$) is the simplest pattern that can translate $\pm1/2$ defects in different directions (see SI for details). The simplest choice for $\alpha$ (centered on the defect) that satisfies the above constraints is
\begin{equation}
    \alpha(\b{r})=\begin{cases}
    \alpha_{0}[1+A\sin(2\phi)+B\cos(2\phi)]\quad (r\leq R)\\
    0\quad(r>R)
    \end{cases}\;,\label{eq:tweezer}
\end{equation}
where $\phi$ is the polar angle about the defect, $\alpha_0<0$ is overall (extensile) activity and different choices of $A,B$ (with $\sqrt{A^2+B^2}\leq 1$) dictate different flow patterns for both $\pm 1/2$ defects (see SI and Figs.~S1-S2 for details). 
}

 We illustrate this idea by using an active tweezer with a time-varying quadrupolar activity profile to move a $-1/2$ defect along a bent trajectory (Fig.~\ref{fig2}A-C, Movie~S1). A $90^\circ$ rotation of the profile causes the $-1/2$ defect to move in an orthogonal direction. A $+1/2$ defect can also be moved along a similar trajectory using a different active tweezer profile (Fig.~\ref{fig2}D-F, Movie~S2). In both cases, the defect successfully tracks the motion of the tweezer (Fig.~\ref{fig2}C and F) for a range of disc speeds $V$ that are comparable or smaller than the activity induced speed $|\b{u}_0|$, i.e., $V\leq |\b{u}_0|$, but very rapid disc motion ($V\gg |\b{u}_0|$) leaves the defect behind, see SI for details on tweezer protocols (Tables~S1-S5) and their characterization (Figs.~S3-S4).

More complex defect manipulations are also possible. As an example, we implement an active tweezer based protocol for simultaneous multidefect control and use it to accomplish a nontrivial defect exchange and braiding task (Fig.~\ref{fig2}G-I, Movie~S3). In a uniformly ordered nematic, we nucleate two pairs of $\pm 1/2$ defects using a high activity ramp within a localized elliptic patch that controls the initial orientation of the defect pair created (Fig.~\ref{fig2}G, right). After the defect pairs are separated, the $-1/2$ defects are braided around each other using active tweezers and finally annihilated with the $+1/2$ defect from the opposite pair, accomplishing pair exchange (Fig.~\ref{fig2}G, I; Movie S3). Although defects also experience elastic forces due to distortions of the nematic \cite{chaikin2000principles}, the local active forces are sufficiently strong to overcome any elastic interaction.
The world lines of the four defects, from creation to annihilation, trace out a closed braid with four crossings forming the figure eight ($4_1$) knot in space-time (Fig.~\ref{fig2}H), which is the simplest, yet nontrivial, achiral prime knot. Notably, unlike active nematics with homogeneous activity, where the spontaneous motility of $+1/2$ defects alone drives autonomous braiding of defect trajectories \cite{tan2019topological}, our example in Fig.~\ref{fig2}G-I demonstrates braiding of the usually disregarded $-1/2$ defect.

Active tweezers hence generate patterned flows that enable control of arbitrarily complex braided and knotted trajectories for both $\pm 1/2$ defects. These capabilities suggest potential strategies for controlling local fluid mixing \cite{tan2019topological,serra2023defect} and provide key steps towards developing reconfigurable space-time assemblies of active defects for programmable logic devices \cite{zhang2022logic,kos2022nematic}. 

\section*{Controlling interacting defect collectives}
\subsection*{Active defect hydrodynamics}
Having demonstrated the capability for controlling individual or a few defects, we extend our framework to address multidefect control at the collective level. To do so, we need to account for two additional features. The first is elastic forces between defects that take the form of Coulomb interactions, familiar from passive liquid crystal physics \cite{chaikin2000principles}. The second is the well known propensity of active nematics to develop chaotic flows and active turbulence (for sufficiently high activity, $|\alpha|\gg K\Gamma/\gamma$), accompanied by swirling $\pm 1/2$ defect pairs that spontaneously unbind and proliferate \cite{doostmohammadi2018active,shankar2022topological}. In the regime of many such unbound defects, a coarse-grained hydrodynamic approach for the defect gas is warranted. Following previous work by some of us \cite{shankar2018defect,shankar2019hydrodynamics}, we develop effective defect hydrodynamic equations that average over fluctuations on the scale of the mean defect spacing and the defect lifetime to describe the distribution of $\pm1/2$ defects in terms of smoothly varying fields such as their respective densities $\rho_{\pm}(\b{r},t)=\langle\sum_i\delta[\b{r}-\b{r}^{\pm}_i(t)]\rangle$ ($\b{r}^{\pm}_i$ is the position of the $i$th $\pm1/2$ defect; see SI for details).

A key advance over Ref.~\cite{shankar2019hydrodynamics} is the inclusion of both a polarization field $\b{p}(\b{r},t)=\langle\sum_i\hat{\b{e}}_i\delta[\b{r}-\b{r}^{+}_i(t)]\rangle$ and a complex triatic order parameter $T_3(\b{r},t)=\langle\sum_i\Theta_3^i\delta[\b{r}-\b{r}^{-}_i(t)]\rangle$ to capture collective orientational ordering of $\pm1/2$ defects due to spatial activity gradients (see SI for details). We derive defect hydrodynamics by coarse-graining effective active particle-like dynamics for $\pm1/2$ defects (see SI for details) that combine activity gradient induced motility and rotations from Eqs.~\ref{uw1},~\ref{uw2} with passive elastic interactions and active collective torques (similar to previously derived torques in Refs.~\cite{shankar2018defect,angheluta2021role}). Defect interaction forces and torques are mediated by nematic distortions quantified by the smoothed phase gradient $\b{v}_n=\langle\vec{\del}\theta\rangle$, where $\theta$ is the local nematic orientation (see SI for details).
\begin{figure*}[t]
    \centering{\includegraphics[width=0.7\textwidth]{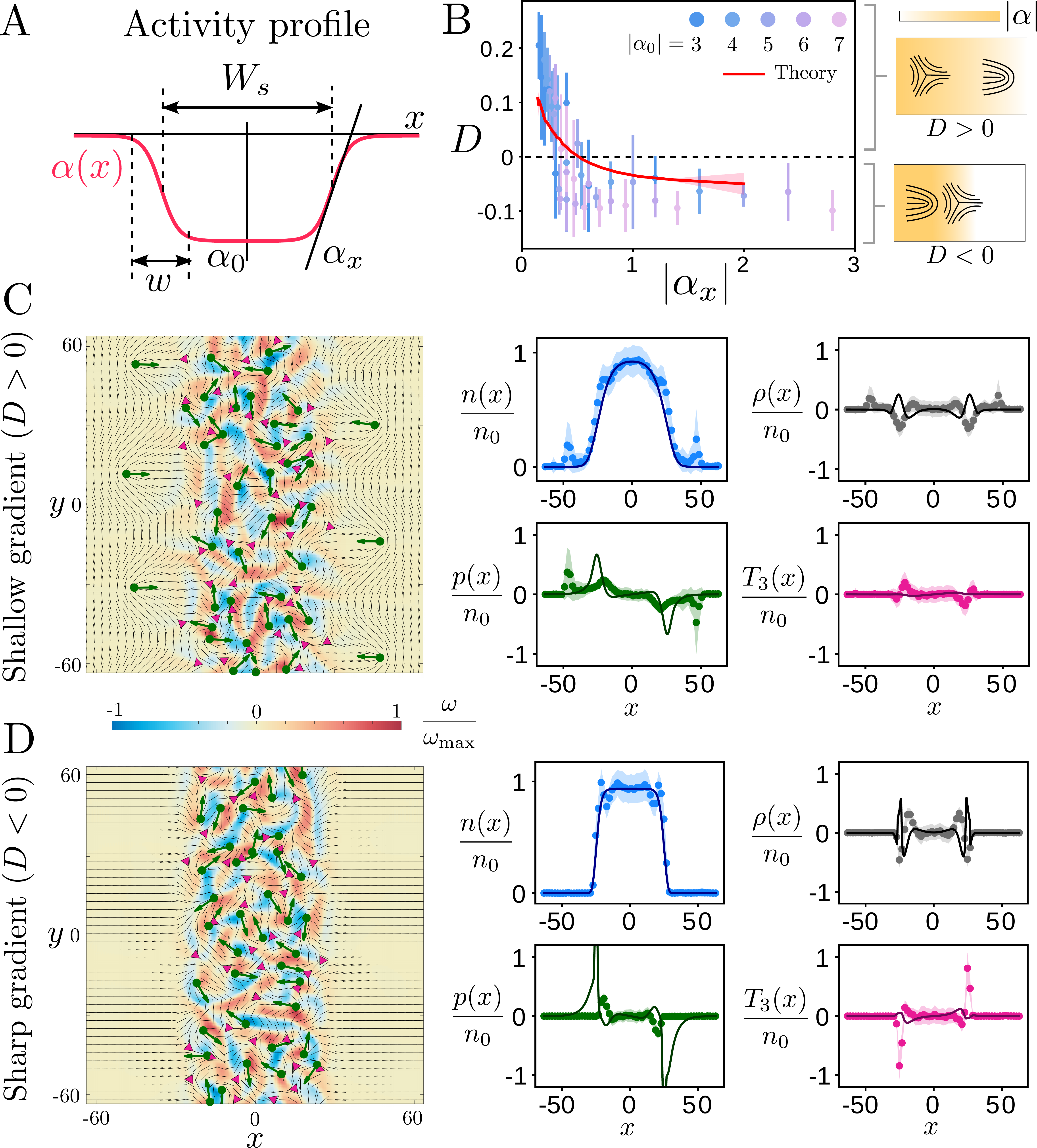}}
    \caption{{\bf Collective spatial patterning of defects realizes charge dipole inversion.} (A) A 1D active strip of width $W_s=50$ with an extensile activity profile $\alpha(x)<0$ that smoothly connects the maximal activity $\alpha_0$ in the interior of the strip to zero outside. In the interfacial region, the activity varies as a sigmoid over a width $w$, generating an activity gradient $\alpha_x\sim\alpha_0/w$. (B) Topological dipole moment ($D$) quantifies interfacial charge separation as a function of activity gradient ($\alpha_x$) for different $\alpha_0$, $w$ (dots: numerical simulation, error bar is one standard deviation). The model (Eqs.~\ref{p}-\ref{j}) quantitatively predicts the dipole flip transition (red line using $|\alpha_0|=5$, shaded region is one standard deviation using an $N=10$ bootstrap subsample; see SI for details). (C-D) The dynamical steady state for shallow ($w=40$, C) and sharp ($w=15$, D) activity gradients (both with $|\alpha_0|=5$) with snapshots shown (left) and spatial defect distributions quantified (right) comparing simulations (dots, shaded region is one standard deviation) with theory (lines, see SI for details). Statistical variation in fitting of model parameters is described in the SI and Figs.~S5 and S8.
    }
    \label{fig3}
\end{figure*}

For a static activity profile $\alpha(x)$ varying in 1D, at steady-state we can set $\rho_\pm=\rho_\pm(x)$, $\b{p}=p(x)\hat{\b{x}}$, $\b{v}_n=v_n(x)\hat{\b{y}}$ etc.~(see SI for details). Neglecting any nonlinear buildup of defect order and expanding to lowest order in gradients, the hydrodynamic equations governing collective flux balance for the $\pm1/2$ defect orientations and their velocities then reduce to (see SI for details)
\begin{align}
    &-\dfrac{1}{\tau_R}p-\dfrac{1}{2}\partial_x(\rho_+\mathcal{V}_+)-\dfrac{1}{2}\rho_+\Omega_++\mu_R\mathcal{V}_+\rho_+\;v_n=0\;,\label{p}\\
    &-\dfrac{1}{\tau_R}T_3-\dfrac{1}{2}\partial_x(\rho_-\mathcal{V}_-)+\mu_R\mathcal{V}_-\rho_-\;v_n=0\;,\label{T3}\\
    &p\mathcal{V}_+-T_3\mathcal{V}_-+2\mu Kn\;v_n=0\;,\label{j}
\end{align}
where the 1D response coefficients $\mathcal{V}_{\pm}(x)$, $\Omega_+(x)$ (Eqs.~\ref{uw1},~\ref{uw2}) are spatial functions of activity (see SI for details), $\tau_R$ is a defect reorientation time
and $\mu_R$ (dimensionless) and $\mu\propto 1/\gamma$ are rotational and translational defect mobilities respectively (see SI for details). The average defect density $n=(\rho_++\rho_-)/2$ is controlled by a steady balance of defect creation and annihilation with $n(x)\propto|\alpha(x)|$ \cite{giomi2015geometry} (see SI for details) and the charge density $\rho=(\rho_+-\rho_-)/2$ is simply obtained from the conservation of topological charge via Gauss' law: $2\pi\rho(x)=\partial_xv_n(x)$ \cite{shankar2019hydrodynamics}. Note, Eq.~\ref{j} balances the motility ($\mathcal{V}_\pm$) of both $\pm1/2$ defects with elastic forces ($\sim \mu Kn v_n$) to set the topological charge current to zero in steady state. Eqs.~\ref{p}-\ref{j} then allow us to compute the spatial distribution of defects in a given 1D activity profile.

\begin{figure*}[t]
    \centering{\includegraphics[width=0.65\textwidth]{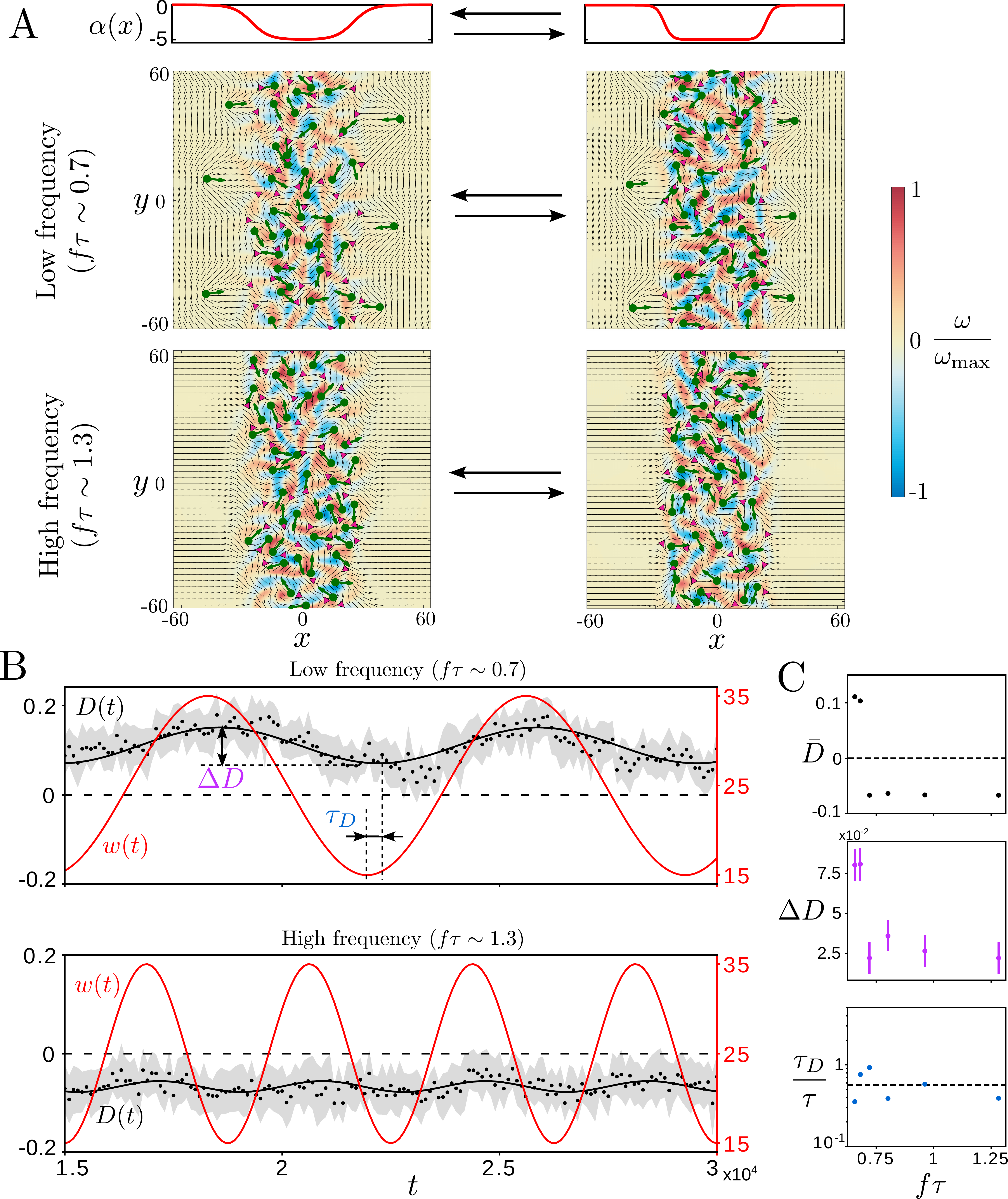}}
    \caption{{\bf Dynamic response of active defects to oscillatory gradients.} (A) The interfacial width ($w$) is varied sinusoidally and its oscillation frequency ($f$) controls defect organization. (B) The dynamical steady state is quantified by a periodic dipole moment $D(t)$ (dots: simulation with shaded region as one standard deviation, black line: sinusoidal fit) which switches from overall $D(t)>0$ at low frequencies ($f\tau\sim 0.7$, top) to $D(t)<0$ at higher frequencies ($f\tau\sim 1.3$, bottom). The time trace of the imposed interfacial width ($w(t)$) is shown in red. (C) The sinusoidal fit (with frequency $f$) of the time varying dipole moment shows how the average value ($\bar{D}$, top), amplitude of oscillation ($\Delta D$, middle) and time delay ($\tau_D$, bottom) vary as a function of drive frequency ($f$). Error bars from the sinusoidal fit correspond to one standard deviation and are smaller than the marker size when not visible.  
    }
    \label{fig4}
\end{figure*}

\subsection*{Static response and dipole inversion}
To validate and test our theoretical predictions, we perform numerical simulations of the full nematodynamic model (Eqs.~\ref{Q},~\ref{u}) using different activity patterns. For simple activity profiles such as a linear or quadratic gradient, we can analytically solve Eqs.~\ref{p}-\ref{j} and obtain quantitative fits for $n(x)$, $\rho(x)$, $p(x)$, $T_3(x)$ (see SI and Figs.~S5-S8 for details). With the phenomenological parameters of our defect hydrodynamic model (Eqs.~\ref{p}-\ref{j}) in hand, we now challenge our model using a more complex activity profile consisting of an active strip of width $W_s=50$ flanked by passive regions (Fig.~\ref{fig3}A). Within the active strip, the maximal activity ($\alpha_0$) is always chosen to be well in the regime of active turbulence. Near the interface, the activity varies in a sigmoidal fashion allowing us to tune the interfacial activity gradient ($\alpha_x\equiv\dd\alpha/\dd x|_{W_s/2}\sim\alpha_0/w$) and the width of the interface ($w$) independently (Fig.~\ref{fig3}A).

For shallow gradients ($|\alpha_x|\leq0.2$), the motile $+1/2$ defects escape and accumulate out of the strip, leaving behind $-1/2$ defects on the inside (Fig.~\ref{fig3}C, left; Movie S4). As a result, the interface develops a topological charge dipole as the activity gradient behaves as an ``electric field'' separating defects by topological charge. This effect, first predicted in Ref.~\cite{shankar2019hydrodynamics}, simply relies on the fact that $+1/2$ defects behave like active particles, which accumulate where they move slowly, and disregards the negligible propulsion of $-1/2$ defects in weak gradients ($\mathcal{V}_-\ll\mathcal{V}_+$, see SI for details). Remarkably, for sharp activity gradients ($|\alpha_x|\geq0.2$), we find a different behavior, wherein the $+1/2$ defects are no longer able to tunnel through the interface, and $-1/2$ defects instead accumulate near the edge of the strip (Fig.~\ref{fig3}D left; Movie S5). Similar effects have been noted previously for individual defects \cite{mozaffari2021defect,zhang2021spatiotemporal}, but not at the collective level.

To quantify the charge separation of defects, we compute a steady state dipole moment $D=(1/2)\int\dd x\;|x|\rho(x)$ and plot it as a function of the interfacial activity gradient $|\alpha_x|$ (Fig.~\ref{fig3}B). Upon varying both the maximal activity ($|\alpha_0|$) and the interfacial width ($w$), we find a common trend, i.e., when the activity gradient $|\alpha_x|$ is increased, the dipole moment switches from $D>0$ (excess $+1/2$ defects on the low activity side) to $D<0$ (excess $+1/2$ defects on the high activity side), see Fig.~\ref{fig3}B. Note that, as $|\alpha_0|\gg 1$ (above the active turbulence threshold), the dipole moment remains nonvanishing even for arbitrarily weak gradients ($|\alpha_x|\to 0$, $w\to\infty$). Following the electrostatic analogy, the active nematic behaves as an unusual polarizable medium with a nonlinear response such that for shallow activity gradients (`weak field') the system behaves as a conventional dielectric, whereas for sharp activity gradients (`strong field'), the system displays a negative static susceptibility, a feature forbidden at equilibrium \cite{landau2013electrodynamics}.

How can we understand the inversion of the dipole moment?
While previous simpler treatments are unable to predict this phenomenon \cite{shankar2019hydrodynamics}, by accounting for the active propulsion and rotation of both $\pm1/2$ defects, our improved defect hydrodynamic model (Eqs.~\ref{p}-\ref{j}) quantitatively captures this dipole flip transition without any fitting parameters (Fig.~\ref{fig3}B, red line using $|\alpha_0|=5$; see SI for details). We next compare the numerically measured spatial distributions of defect density ($n$), charge density ($\rho$) and orientational order parameters ($p$, $T_3$) with our model predictions, with qualitatively comparable results overall (Fig.~\ref{fig3}C-D, right).
The average defect density $n(x)$ is well predicted in both cases ($n_0$ is the maximal defect density in the center) and features of charge separation ($\rho(x)$) and $+1/2$ polarization ($p(x)$) are qualitatively captured for shallow gradients (Fig.~\ref{fig3}C, right; see SI for details). But the model predictions for spatial profiles are less accurate for sharp gradients (Fig.~\ref{fig3}D, right), where $p(x)$ is overestimated and $T_3(x)$ is underestimated (see SI for details).
We attribute the lack of quantitative accuracy in the spatial profile predictions to the neglect of higher gradient and nonlinear terms that affect the defect density and cause saturation of defect order, particularly for strong activity gradients (see SI for details).

Nonetheless, some general features are apparent. Intuitively, the sharp activity interface acts like a virtual wall that blocks the motion of defects through it, as noted previously in simulations \cite{mozaffari2021defect} and experiments \cite{zhang2021spatiotemporal,thijssen2021submersed}. In a thin boundary layer near the edge of the active strip, the locally strong activity gradient causes $-1/2$ defects to self-propel faster than $+1/2$ defects ($\mathcal{V}_-\geq\mathcal{V}_+$), but both $\pm1/2$ defects rapidly lose their motility upon crossing the interface. Similar boundary layers have been recently observed at confining walls as well \cite{hardouin2022active}. As a result, $-1/2$ defects get preferentially trapped along the interface, preventing any $+1/2$ defects from crossing the interface, thus realizing the inverted dipole state.

\begin{figure*}[t]
    \centering{\includegraphics[width=0.7\textwidth]{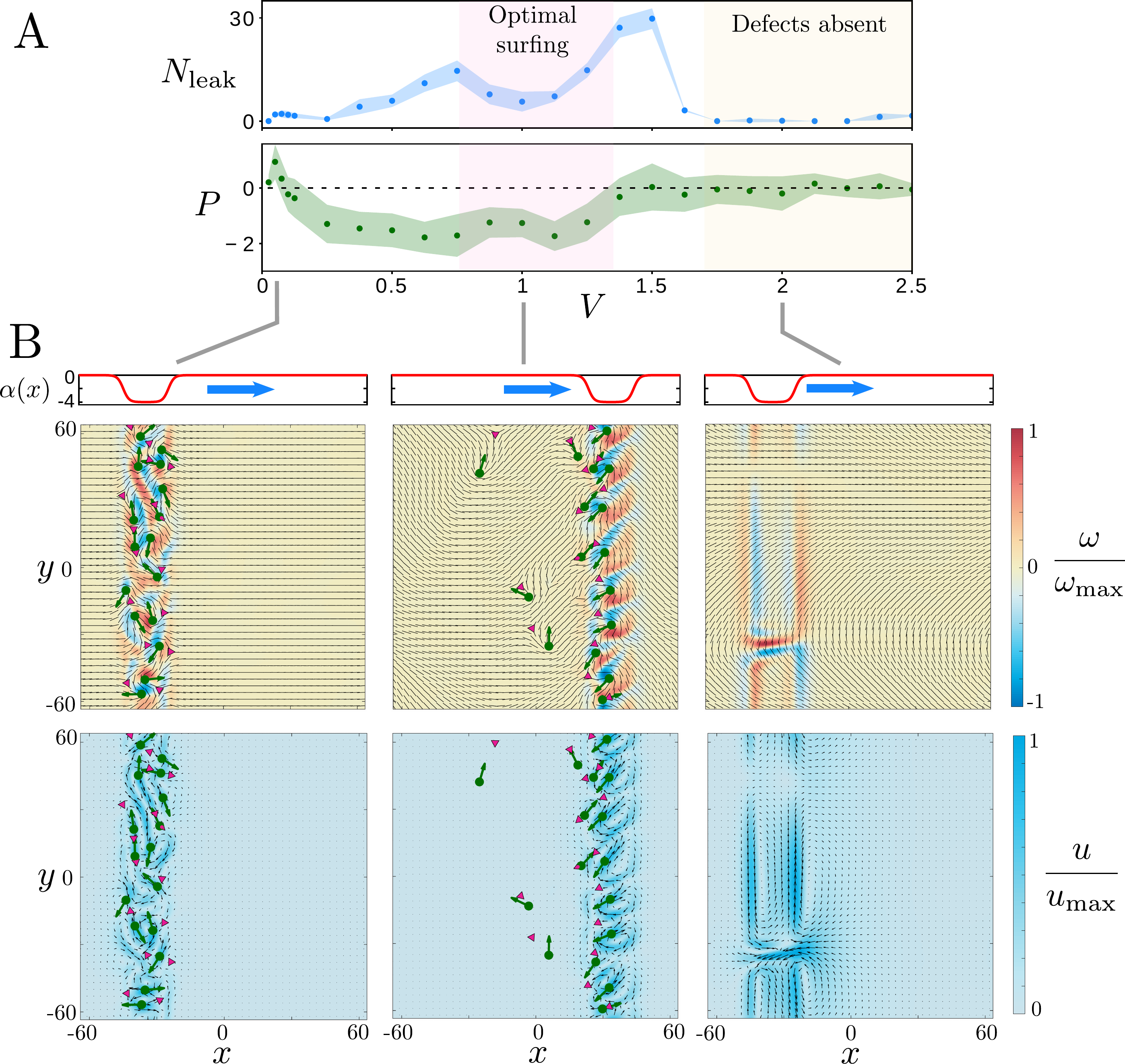}}
    \caption{{\bf Collective transport of active defects by `surfing'.} (A) A 1D active strip ($|\alpha_0|=4$, $W_s=20$, $w=10$) moving with speed $V$ can be used to collectively transport active defects. Transport efficacy is quantified by the total number of defects that leak out of the strip ($N_{\rm leak}$, blue dots) and the total polarization of $+1/2$ defects ($P$, green dots). The blue and green shaded regions represent one standard deviation. Few defects are lost for $V\ll 1$ but the transport is slow, whereas for very high speeds ($V\gg 1$, yellow shaded region), the strip moves faster than defects can nucleate. Only at intermediate speeds ($V\sim 1$, shaded red region) do we obtain optimal defect surfing, where few defects are left behind and the $+1/2$ defects develop significant collective polarization ($|P|\gg 1$). (B) Snapshots show the three characteristic regimes for $V\ll 1$ (left), $V\sim 1$ (middle) and $V\gg 1$ (right) with both vorticity ($\omega$) and flow speed ($u=|\b{u}|$) plotted. 
    }
    \label{fig5}
\end{figure*}

The resulting charge dipole is accompanied by a interfacial defect ordering and a change in the nematic orientation outside the active strip as well. As shown in Fig.~\ref{fig3}C-D, while shallow activity gradients cause the nematic to orient parallel to the interface (due to the escaped $+1/2$ defects whose polarization points down activity gradients), a sharp interface forces the nematic to reorient perpendicular to the interface (due to ordered $-1/2$ defects). Our results demonstrate that defect decorated activity interfaces display tunable `active anchoring' \cite{blow2014biphasic} that allow control of collective defect organization and nematic orientation simply via the strength of the local activity gradient.

\subsection*{Dynamic response and rectification}
We now employ the active strip as a defect patterning motif in a dynamic setting. As shallow interfaces are leaky to the escape of $+1/2$ defects, but sharp interfaces are not, we ask if periodically oscillating the interface width allows us to dynamically control the organization of active defects. We fix $|\alpha_0|=5$ and vary the interface width sinusoidally in time (with frequency $f$) between values $w_\rm{min}=15$ and $w_{\rm{max}}=35$ (Fig.~\ref{fig4}A), so that the average width ($\bar{w}=25$) corresponds to an almost vanishing dipole moment ($D\approx 0$) in the static limit (see Fig.~\ref{fig3}B). The average time $\tau=\bar{w}^2/D_a$ 
it takes $+1/2$ defects to cross the interface with an active diffusion constant $D_a=\langle|\b{u}_0|^2\rangle\tau_R/2\sim 0.82$
(see SI for details) sets a characteristic time scale that controls the dynamic response. At low oscillation frequencies ($f\tau\sim0.7$, with $\tau\sim765$), $+1/2$ defects have sufficient time to escape through the shallow interface and remain trapped in the passive region when the interface becomes sharp (Fig.~\ref{fig4}A, Movie S6). As a result, the system develops a steady-state dipole moment that is always positive ($D(t)>0$) and oscillates in sync with the interface (Fig.~\ref{fig4}B, top), albeit with a time delay ($\tau_D$, Fig.~\ref{fig4}B-C).

For a higher driving frequency ($f\tau\sim 1.3$), $+1/2$ defects have insufficient time to escape and they get dynamically trapped to the active strip (Fig.~\ref{fig4}A, Movie S7). The system then \emph{dynamically} realizes an inverted dipole state with $D(t)<0$ (Fig.~\ref{fig4}B, bottom). We quantify this dynamic transition by performing a sinusoidal fit of the steady-state dipole moment and plot the time-averaged dipole moment ($\bar{D}$), the amplitude of oscillations ($\Delta D$) and the time delay in the response ($\tau_D$) as a function of the driving frequency $f$ (Fig.~\ref{fig4}C). While $\bar{D}$ and $\Delta D$ show stronger variation upon increasing frequency (with $\bar{D}$ switching sign), the average time delay $\tau_D\sim 0.6\tau$ is intrinsic to the active defect gas and does not vary systematically with the drive (Fig.~\ref{fig4}C).

\subsection*{Collective transport and optimal surfing}
Along with defect patterning, another basic task is defect transport. As a simple example, we use a 1D moving active strip (with speed $V$) to illustrate how active defects can be collectively transported in space. For simplicity, we fix the maximal activity $|\alpha_0|=4$ and choose the strip geometry to have sharp interfaces ($W_s=20$, $w=10$) to ensure defects are trapped to the active region when the strip is stationary. To achieve rapid and efficient transport, it is natural to consider large $V$, but fast motion of the active strip can cause defects to be left behind, despite the sharp interfaces. To quantify this trade-off we compute (in steady-state) the total number of defects that escape and leak out of the active strip ($N_\rm{leak}$, Fig.~\ref{fig5}A) and the net horizontal polarization of all the $+1/2$ defects ($P=\int\dd\b{r}\;p$, Fig.~\ref{fig5}A).

For slow speeds ($V\ll 1$) the transport time is long but the defects remain largely localized to the active region ($N_{\rm{leak}}\sim 0$) with no net polarization ($P\sim 0$) as they follow the travelling strip adiabatically (Fig.~\ref{fig5}B left, Movie S8). For very rapid motion of the strip ($V\gg 1$, shaded yellow region in Fig.~\ref{fig5}A), there is insufficient time to nucleate and carry defects (Fig.~\ref{fig5}B right; Movie S10). Only for intermediate speeds ($V\sim 1$, shaded red region in Fig.~\ref{fig5}A) do we obtain optimal collective transport of defects characterized by an enhanced carrying capacity (decrease in $N_{\rm{leak}}$, Fig.~\ref{fig5}A) and significant $+1/2$ polarization ($|P|\gg 1$, Fig.~\ref{fig5}A). In this regime, the active strip moves at a speed comparable to the activity induced propulsion of the defects, allowing the defects to collectively `surf' the imposed travelling wave of activity, and organize the otherwise turbulent interior of the active strip into a state with spatially patterned flow and vortices (Fig.~\ref{fig5}B middle, Movie S9).

\section*{Discussion and Conclusion}
In this work, we have demonstrated how topological defects offer robust particle-like excitations that can be functionally controlled to manipulate continuous active fluids in space and time. Our symmetry based additive framework enables the construction of active topological tweezers for basic defect operations that provide the basis of complex computation and logic in a fluidic system \cite{kos2022nematic,zhang2022logic}. By extending the framework to incorporate defect interactions, we developed a coarse-grained hydrodynamic description of active defects at the collective level, enabling the characterization of large-scale patterning and dynamics of the defect gas. Activity gradients are shown to behave as ``electric fields'' that segregate defects by charge and topologically polarize the active fluid, albeit with unusual responses that mimic certain dielectric metamaterials. Complementing the spatial response, we also probe the dynamic response of defects and highlight simple strategies for patterning and transporting large collections of active defects.

Our work provides a general framework to control and rationally design active materials as \emph{meta}fluids for soft microrobotics \cite{nitta2021printable,jia20223d} and defect based soft logic \cite{kos2022nematic,zhang2022logic}. Current experiments on light-controlled active fluids \cite{ross2019controlling,zhang2021spatiotemporal,linnea2022} offer a natural platform to deploy our control strategies. Beyond engineered systems, active defects have been identified in biological tissues and cellular monolayers, and proposed to act as sites of biological function and morphogenesis \cite{saw2017topological,kawaguchi2017topological,copenhagen2021topological,maroudas2021topological}. Our work suggests that a similar symmetry based approach can be used to understand how active defects get functionalized in biological systems, paving the way for controlling and designing defect based autonomous materials from living matter with adaptive and programmable functionality.

\setcounter{figure}{0}
\renewcommand{\figurename}{Extended Data Figure}

\matmethods{
\label{Method}Extended data on the details of the numerical modeling and theoretical calculations. Further details on model fitting and analysis are provided in the SI.
\subsection*{Data and code availability}
Code used for the numerical simulations is available on \href{github.com/LVDScharrer/ATC}{github.com/LVDScharrer/ATC}. All other data needed to reproduce the results in this paper are provided in the Methods and Supplementary Information.
\subsection*{Active nematodynamics} 
\label{sec:nematic}
The orientational order of the active nematic in 2D is locally characterized by an alignment tensor $Q_{ij}=S(\hat{n}_i\hat{n}_j-\delta_{ij}/2)$ with the director $\hat{\b{n}}=(\cos\theta,\sin\theta)$, where $S$ is the scalar order parameter and $\theta$ refers to the director angle.
A continuum hydrodynamic description of the 2D active nematic film is given by (Eqs.~\ref{Q},~\ref{u}) 
\begin{align}
    &\partial_t\b{Q}+\b{u}\cdot\vec{\del}\b{Q}=\b{S}(\b{u},\b{Q})+\dfrac{1}{\gamma}\b{H}\;,\label{Q1}\\
    &-\Gamma\b{u}+\eta\del^2\b{u}-\vec{\del}\Pi+\vec{\del}\cdot\left(\vec{\sigma}^a+\vec{\sigma}^{el}\right)=\b{0}\;,
\label{u1}
\end{align}
with the molecular field $\b{H}=(a_2-a_4\tr[\b{Q}^2])\b{Q}+K\del^2\b{Q}$ ($a_{2,4}>0$). Note here we have already assumed the system prefers an ordered nematic phase in the absence of activity. Without loss of generality, we set $a_4=a_2$. Then the equilibrium ground state is an ordered homogeneous nematic with $S=S_0=2$. The flow coupling $\b{S}(\b{u},\b{Q})$ is given by \cite{beris1994thermodynamics}
\begin{align}
     S_{ij}&=Q_{ik}W_{kj}-W_{ik}Q_{kj}- 2\lambda Q_{k\ell}E_{k\ell}Q_{ij}\nonumber\\
     &\quad+\lambda E_{ij} + \lambda( E_{ik}Q_{kj} +Q_{ik}E_{kj}-\delta_{ij}Q_{k\ell}E_{k\ell} )\;,
\end{align}
where $\lambda$ is the flow-alignment parameter ($|\lambda|>1$ for flow-aligning systems), $E_{ij} = (\partial_i u_j + \partial_j u_i)/2$ is the strain-rate tensor (purely deviatoric as $\vec{\del}\cdot\b{u}=0$), and $W_{ij} = (\partial_i u_j - \partial_j u_i)/2$ is the vorticity tensor.

The active stress is $\vec{\sigma}^a=\alpha\b{Q}$ \cite{marchetti2013hydrodynamics} and the passive liquid-crystal stress is given by
\begin{align}
    \sigma^{el}_{ij}&=Q_{ik}  H_{kj} -  H_{ik}Q_{kj} - K\partial_i Q_{kl} \partial_j Q_{kl}+\lambda(2Q_{ij} +  \delta_{ij}) Q_{kl}  H_{kl}\nonumber\\
    &\quad -\lambda  H_{ik}\left(Q_{kj} +\dfrac{1}{2}\delta_{kj}\right) -\lambda \left(Q_{ik} + \dfrac{1}{2}\delta_{ik}\right)  H_{kj}\;.\label{eq:sigel}
\end{align}

The pressure $\Pi$ enforces incompressibility ($\vec{\del}\cdot\b{u}=0$) and is calculated from the pressure Poisson equation obtained by taking the divergence of Stokes' equation (Eq.~\ref{u1}):
\begin{equation}
    \nabla^2 \Pi = \partial_i \partial_j \left(\sigma^{el}_{ij} + \sigma^a_{ij}  \right)
\end{equation}
We note that only symmetric components of $\vec{\sigma}^{el}$ will contribute to the pressure, due to the symmetry of the $\partial_i\partial_j$ operator.

\subsection*{Numerical simulations}
\label{subsec:num}
In order to numerically simulate the nematodynamic equations, we nondimensionalize Eqs.~\ref{Q1},~\ref{u1}. As mentioned in the main text, we employ the nematic coherence length $\xi = \sqrt{K/a_2}$ as our unit of length, the elastic relaxation time $\tau_n=\gamma/a_2$ as our unit of time, and a passive stress scale $K\Gamma/\gamma$ as our unit of stress. We are then left with four nondimensionless parameters: $\tilde{\ell}_\eta=\ell_\eta/\xi$, $\tilde{\gamma}=\gamma/\eta$, $\tilde{\alpha}=\alpha\gamma/(K\Gamma)$ and $\lambda$. The topological tweezer demonstrations were performed with $\tilde{\ell}_\eta = 5$, chosen to display flow patterns more clearly, while the collective effect simulations were performed at a lower viscosity $\tilde{\ell_\eta} = \sqrt{5}$, to increase defect density and improve the statistics of defect averages. For all simulations, we set $\tilde{\gamma}=0.1$ and $\lambda=1.8$.

Numerical simulations of the continuum nematodynamic equations (Eqs.~\ref{Q1},~\ref{u1}) were performed with a custom Matlab code using a second order pseudospectral time-exponentiation scheme for numerical integration, on a lattice of $256 \times 256$ gridpoints with periodic boundary conditions. We used a grid spacing of $\dd x = 0.5$ and timestep $\dd t = 0.1$. The $-1/2$ and $+1/2$ tweezer demonstrations in Fig.~\ref{fig2} were run for a total of $1,520$ and $1,800$ timesteps respectively, while the braiding procedure was run for $6,500$ timesteps. The dipole moment datapoints reported in Fig.~\ref{fig3} were computed using $N=40$ independent simulations, each running for $10^5$ timesteps. The data for oscillating activity gradients (Fig.~\ref{fig4}) were computed using $N=6$ independent simulations, each run for $3 \times 10^5$ timesteps, and the demonstrations with traveling activity patterns (Fig.~\ref{fig5}) were run for $5 \times 10^4$ timesteps. All simulations were performed on an NVIDIA GeForce GTX 1060 Mobile graphics card, and we estimate that each $10^4$ timesteps of simulation time corresponded to a wall-clock runtime of $\sim12$~minutes, resulting in a typical overall runtime of $\sim150$~hours.

}

\showmatmethods{} 

\acknow{SS acknowledges support from the Harvard Society of Fellows. LVDS acknowledges support from the College of Creative Studies' Francesc Roig Summer Undergraduate Research Fund. This research was supported in part by the National Science Foundation under Grants No.~PHY-1748958 (MJB and LS) and DMR-2041459 (MCM and LS).
}

\showacknow{} 


\begin{thebibliography}{10}

\bibitem{xia2022responsive}
X Xia, CM Spadaccini, JR Greer, Responsive materials architected in space and
  time.
\newblock {\em\protect\JournalTitle{Nature Reviews Materials}} \textbf{7},
  1--19 (2022).

\bibitem{jacobs2016self}
WM Jacobs, D Frenkel, Self-assembly of structures with addressable complexity.
\newblock {\em\protect\JournalTitle{Journal of the American Chemical Society}}
  \textbf{138}, 2457--2467 (2016).

\bibitem{dey2021dna}
S Dey, et~al., Dna origami.
\newblock {\em\protect\JournalTitle{Nature Reviews Methods Primers}}
  \textbf{1}, 1--24 (2021).

\bibitem{zeravcic2017colloquium}
Z Zeravcic, VN Manoharan, MP Brenner, Colloquium: Toward living matter with
  colloidal particles.
\newblock {\em\protect\JournalTitle{Reviews of Modern Physics}} \textbf{89},
  031001 (2017).

\bibitem{bertoldi2017flexible}
K Bertoldi, V Vitelli, J Christensen, M Van~Hecke, Flexible mechanical
  metamaterials.
\newblock {\em\protect\JournalTitle{Nature Reviews Materials}} \textbf{2},
  1--11 (2017).

\bibitem{zangeneh2021analogue}
F Zangeneh-Nejad, DL Sounas, A Al{\`u}, R Fleury, Analogue computing with
  metamaterials.
\newblock {\em\protect\JournalTitle{Nature Reviews Materials}} \textbf{6},
  207--225 (2021).

\bibitem{yasuda2021mechanical}
H Yasuda, et~al., Mechanical computing.
\newblock {\em\protect\JournalTitle{Nature}} \textbf{598}, 39--48 (2021).

\bibitem{chaikin2000principles}
PM Chaikin, TC Lubensky, {\em Principles of condensed matter physics}.
\newblock (Cambridge university press), (2000).

\bibitem{romming2013writing}
N Romming, et~al., Writing and deleting single magnetic skyrmions.
\newblock {\em\protect\JournalTitle{Science}} \textbf{341}, 636--639 (2013).

\bibitem{fert2017magnetic}
A Fert, N Reyren, V Cros, Magnetic skyrmions: advances in physics and potential
  applications.
\newblock {\em\protect\JournalTitle{Nature Reviews Materials}} \textbf{2},
  1--15 (2017).

\bibitem{berciu2005manipulating}
M Berciu, TG Rappoport, B Jank{\'o}, Manipulating spin and charge in magnetic
  semiconductors using superconducting vortices.
\newblock {\em\protect\JournalTitle{Nature}} \textbf{435}, 71--75 (2005).

\bibitem{ge2016nanoscale}
JY Ge, et~al., Nanoscale assembly of superconducting vortices with scanning
  tunnelling microscope tip.
\newblock {\em\protect\JournalTitle{Nature communications}} \textbf{7}, 1--7
  (2016).

\bibitem{tkalec2011reconfigurable}
U Tkalec, M Ravnik, S {\v{C}}opar, S {\v{Z}}umer, I Mu{\v{s}}evi{\v{c}},
  Reconfigurable knots and links in chiral nematic colloids.
\newblock {\em\protect\JournalTitle{Science}} \textbf{333}, 62--65 (2011).

\bibitem{tai2019three}
JSB Tai, II Smalyukh, Three-dimensional crystals of adaptive knots.
\newblock {\em\protect\JournalTitle{Science}} \textbf{365}, 1449--1453 (2019).

\bibitem{foster2019two}
D Foster, et~al., Two-dimensional skyrmion bags in liquid crystals and
  ferromagnets.
\newblock {\em\protect\JournalTitle{Nature Physics}} \textbf{15}, 655--659
  (2019).

\bibitem{marchetti2013hydrodynamics}
MC Marchetti, et~al., Hydrodynamics of soft active matter.
\newblock {\em\protect\JournalTitle{Reviews of modern physics}} \textbf{85},
  1143 (2013).

\bibitem{doostmohammadi2018active}
A Doostmohammadi, J Ign{\'e}s-Mullol, JM Yeomans, F Sagu{\'e}s, Active
  nematics.
\newblock {\em\protect\JournalTitle{Nature communications}} \textbf{9}, 1--13
  (2018).

\bibitem{shankar2022topological}
S Shankar, A Souslov, MJ Bowick, MC Marchetti, V Vitelli, Topological active
  matter.
\newblock {\em\protect\JournalTitle{Nature Reviews Physics}} \textbf{4},
  380--398 (2022).

\bibitem{zhou2014living}
S Zhou, A Sokolov, OD Lavrentovich, IS Aranson, Living liquid crystals.
\newblock {\em\protect\JournalTitle{Proceedings of the National Academy of
  Sciences}} \textbf{111}, 1265--1270 (2014).

\bibitem{li2019data}
H Li, et~al., Data-driven quantitative modeling of bacterial active nematics.
\newblock {\em\protect\JournalTitle{Proceedings of the National Academy of
  Sciences}} \textbf{116}, 777--785 (2019).

\bibitem{sanchez2012spontaneous}
T Sanchez, DT Chen, SJ DeCamp, M Heymann, Z Dogic, Spontaneous motion in
  hierarchically assembled active matter.
\newblock {\em\protect\JournalTitle{Nature}} \textbf{491}, 431--434 (2012).

\bibitem{bricard2013emergence}
A Bricard, JB Caussin, N Desreumaux, O Dauchot, D Bartolo, Emergence of
  macroscopic directed motion in populations of motile colloids.
\newblock {\em\protect\JournalTitle{Nature}} \textbf{503}, 95--98 (2013).

\bibitem{duclos2018spontaneous}
G Duclos, et~al., Spontaneous shear flow in confined cellular nematics.
\newblock {\em\protect\JournalTitle{Nature physics}} \textbf{14}, 728--732
  (2018).

\bibitem{giomi2013defect}
L Giomi, MJ Bowick, X Ma, MC Marchetti, Defect annihilation and proliferation
  in active nematics.
\newblock {\em\protect\JournalTitle{Physical review letters}} \textbf{110},
  228101 (2013).

\bibitem{shankar2018defect}
S Shankar, S Ramaswamy, MC Marchetti, MJ Bowick, Defect unbinding in active
  nematics.
\newblock {\em\protect\JournalTitle{Physical review letters}} \textbf{121},
  108002 (2018).

\bibitem{tan2019topological}
AJ Tan, et~al., Topological chaos in active nematics.
\newblock {\em\protect\JournalTitle{Nature Physics}} \textbf{15}, 1033--1039
  (2019).

\bibitem{alert2022active}
R Alert, J Casademunt, JF Joanny, Active turbulence.
\newblock {\em\protect\JournalTitle{Annual Review of Condensed Matter Physics}}
  \textbf{13} (2022).

\bibitem{serra2023defect}
M Serra, L Lemma, L Giomi, Z Dogic, L Mahadevan, Defect-mediated dynamics of
  coherent structures in active nematics.
\newblock {\em\protect\JournalTitle{Nature Physics}} \textbf{19}, 1--7 (2023).

\bibitem{needleman2017active}
D Needleman, Z Dogic, Active matter at the interface between materials science
  and cell biology.
\newblock {\em\protect\JournalTitle{Nature Reviews Materials}} \textbf{2},
  1--14 (2017).

\bibitem{zhang2021autonomous}
R Zhang, A Mozaffari, JJ de~Pablo, Autonomous materials systems from active
  liquid crystals.
\newblock {\em\protect\JournalTitle{Nature Reviews Materials}} \textbf{6},
  437--453 (2021).

\bibitem{peng2016command}
C Peng, T Turiv, Y Guo, QH Wei, OD Lavrentovich, Command of active matter by
  topological defects and patterns.
\newblock {\em\protect\JournalTitle{Science}} \textbf{354}, 882--885 (2016).

\bibitem{opathalage2019self}
A Opathalage, et~al., Self-organized dynamics and the transition to turbulence
  of confined active nematics.
\newblock {\em\protect\JournalTitle{Proceedings of the National Academy of
  Sciences}} \textbf{116}, 4788--4797 (2019).

\bibitem{keber2014topology}
FC Keber, et~al., Topology and dynamics of active nematic vesicles.
\newblock {\em\protect\JournalTitle{Science}} \textbf{345}, 1135--1139 (2014).

\bibitem{ellis2018curvature}
PW Ellis, et~al., Curvature-induced defect unbinding and dynamics in active
  nematic toroids.
\newblock {\em\protect\JournalTitle{Nature Physics}} \textbf{14}, 85--90
  (2018).

\bibitem{guillamat2016control}
P Guillamat, J Ign{\'e}s-Mullol, F Sagu{\'e}s, Control of active liquid
  crystals with a magnetic field.
\newblock {\em\protect\JournalTitle{Proceedings of the National Academy of
  Sciences}} \textbf{113}, 5498--5502 (2016).

\bibitem{thijssen2021submersed}
K Thijssen, et~al., Submersed micropatterned structures control active nematic
  flow, topology, and concentration.
\newblock {\em\protect\JournalTitle{Proceedings of the National Academy of
  Sciences}} \textbf{118}, e2106038118 (2021).

\bibitem{dervaux2017light}
J Dervaux, M Capellazzi~Resta, P Brunet, Light-controlled flows in active
  fluids.
\newblock {\em\protect\JournalTitle{Nature Physics}} \textbf{13}, 306--312
  (2017).

\bibitem{arlt2018painting}
J Arlt, VA Martinez, A Dawson, T Pilizota, WC Poon, Painting with light-powered
  bacteria.
\newblock {\em\protect\JournalTitle{Nature communications}} \textbf{9}, 1--7
  (2018).

\bibitem{frangipane2018dynamic}
G Frangipane, et~al., Dynamic density shaping of photokinetic e. coli.
\newblock {\em\protect\JournalTitle{Elife}} \textbf{7}, e36608 (2018).

\bibitem{ross2019controlling}
TD Ross, et~al., Controlling organization and forces in active matter through
  optically defined boundaries.
\newblock {\em\protect\JournalTitle{Nature}} \textbf{572}, 224--229 (2019).

\bibitem{zhang2021spatiotemporal}
R Zhang, et~al., Spatiotemporal control of liquid crystal structure and
  dynamics through activity patterning.
\newblock {\em\protect\JournalTitle{Nature materials}} \textbf{20}, 875--882
  (2021).

\bibitem{linnea2022}
LM Lemma, et~al., Spatiotemporal patterning of extensile active stresses in
  microtubule-based active fluids (2022).

\bibitem{norton2020optimal}
MM Norton, P Grover, MF Hagan, S Fraden, Optimal control of active nematics.
\newblock {\em\protect\JournalTitle{Physical review letters}} \textbf{125},
  178005 (2020).

\bibitem{shankar2022optimal}
S Shankar, V Raju, L Mahadevan, Optimal transport and control of active drops.
\newblock {\em\protect\JournalTitle{Proceedings of the National Academy of
  Sciences}} \textbf{119}, e2121985119 (2022).

\bibitem{falk2021learning}
MJ Falk, V Alizadehyazdi, H Jaeger, A Murugan, Learning to control active
  matter.
\newblock {\em\protect\JournalTitle{Physical Review Research}} \textbf{3},
  033291 (2021).

\bibitem{shankar2019hydrodynamics}
S Shankar, MC Marchetti, Hydrodynamics of active defects: From order to chaos
  to defect ordering.
\newblock {\em\protect\JournalTitle{Physical Review X}} \textbf{9}, 041047
  (2019).

\bibitem{mozaffari2021defect}
A Mozaffari, R Zhang, N Atzin, J de~Pablo, Defect spirograph: Dynamical
  behavior of defects in spatially patterned active nematics.
\newblock {\em\protect\JournalTitle{Physical Review Letters}} \textbf{126},
  227801--227801 (2021).

\bibitem{ruske2022activity}
LJ Ruske, JM Yeomans, Activity gradients in two-and three-dimensional active
  nematics.
\newblock {\em\protect\JournalTitle{Soft Matter}} \textbf{18}, 5654--5661
  (2022).

\bibitem{yang2022}
F Yang, S Liu, HJ Lee, M Thomson, Dynamic flow control through active matter
  programming language (2022).

\bibitem{zhang2022logic}
R Zhang, A Mozaffari, JJ de~Pablo, Logic operations with active topological
  defects.
\newblock {\em\protect\JournalTitle{Science advances}} \textbf{8}, eabg9060
  (2022).

\bibitem{tang2021alignment}
X Tang, JV Selinger, Alignment of a topological defect by an activity gradient.
\newblock {\em\protect\JournalTitle{Physical Review E}} \textbf{103}, 022703
  (2021).

\bibitem{jonas2022}
J R{\o}nning, MC Marchetti, L Angheluta, Defect self-propulsion in active
  nematic films with spatially-varying activity (2022).

\bibitem{pismen2013dynamics}
L Pismen, Dynamics of defects in an active nematic layer.
\newblock {\em\protect\JournalTitle{Physical Review E}} \textbf{88}, 050502
  (2013).

\bibitem{bernshtein1967co}
N Bernshte{\u\i}n, {\em The Co-ordination and Regulation of Movements}.
\newblock (Oxford, Pergamon Press), (1967).

\bibitem{vromans2016orientational}
AJ Vromans, L Giomi, Orientational properties of nematic disclinations.
\newblock {\em\protect\JournalTitle{Soft matter}} \textbf{12}, 6490--6495
  (2016).

\bibitem{tang2017orientation}
X Tang, JV Selinger, Orientation of topological defects in 2d nematic liquid
  crystals.
\newblock {\em\protect\JournalTitle{Soft matter}} \textbf{13}, 5481--5490
  (2017).

\bibitem{irvine2013dislocation}
WT Irvine, AD Hollingsworth, DG Grier, PM Chaikin, Dislocation reactions, grain
  boundaries, and irreversibility in two-dimensional lattices using topological
  tweezers.
\newblock {\em\protect\JournalTitle{Proceedings of the National Academy of
  Sciences}} \textbf{110}, 15544--15548 (2013).

\bibitem{kos2022nematic}
{\v{Z}} Kos, J Dunkel, Nematic bits and universal logic gates.
\newblock {\em\protect\JournalTitle{Science Advances}} \textbf{8}, eabp8371
  (2022).

\bibitem{angheluta2021role}
L Angheluta, Z Chen, MC Marchetti, MJ Bowick, The role of fluid flow in the
  dynamics of active nematic defects.
\newblock {\em\protect\JournalTitle{New Journal of Physics}} \textbf{23},
  033009 (2021).

\bibitem{giomi2015geometry}
L Giomi, Geometry and topology of turbulence in active nematics.
\newblock {\em\protect\JournalTitle{Physical Review X}} \textbf{5}, 031003
  (2015).

\bibitem{landau2013electrodynamics}
LD Landau, et~al., {\em Electrodynamics of continuous media}.
\newblock (elsevier) Vol.{}~8, (2013).

\bibitem{hardouin2022active}
J Hardo{\"u}in, J Laurent, T Lopez-Leon, J Ign{\'e}s-Mullol, F Sagu{\'e}s,
  Active boundary layers in confined active nematics.
\newblock {\em\protect\JournalTitle{Nature communications}} \textbf{13}, 6675
  (2022).

\bibitem{blow2014biphasic}
ML Blow, SP Thampi, JM Yeomans, Biphasic, lyotropic, active nematics.
\newblock {\em\protect\JournalTitle{Physical review letters}} \textbf{113},
  248303 (2014).

\bibitem{nitta2021printable}
T Nitta, Y Wang, Z Du, K Morishima, Y Hiratsuka, A printable active network
  actuator built from an engineered biomolecular motor.
\newblock {\em\protect\JournalTitle{Nature Materials}} \textbf{20}, 1149--1155
  (2021).

\bibitem{jia20223d}
H Jia, et~al., 3d printed protein-based robotic structures actuated by
  molecular motor assemblies.
\newblock {\em\protect\JournalTitle{Nature Materials}} \textbf{21}, 703--709
  (2022).

\bibitem{saw2017topological}
TB Saw, et~al., Topological defects in epithelia govern cell death and
  extrusion.
\newblock {\em\protect\JournalTitle{Nature}} \textbf{544}, 212--216 (2017).

\bibitem{kawaguchi2017topological}
K Kawaguchi, R Kageyama, M Sano, Topological defects control collective
  dynamics in neural progenitor cell cultures.
\newblock {\em\protect\JournalTitle{Nature}} \textbf{545}, 327--331 (2017).

\bibitem{copenhagen2021topological}
K Copenhagen, R Alert, NS Wingreen, JW Shaevitz, Topological defects promote
  layer formation in myxococcus xanthus colonies.
\newblock {\em\protect\JournalTitle{Nature Physics}} \textbf{17}, 211--215
  (2021).

\bibitem{maroudas2021topological}
Y Maroudas-Sacks, et~al., Topological defects in the nematic order of actin
  fibres as organization centres of hydra morphogenesis.
\newblock {\em\protect\JournalTitle{Nature Physics}} \textbf{17}, 251--259
  (2021).

\bibitem{beris1994thermodynamics}
AN Beris, BJ Edwards, {\em Thermodynamics of flowing systems: with internal
  microstructure}.
\newblock (Oxford University Press on Demand) No.{}~36, (1994).

\bibitem{giomi2014defect}
L Giomi, MJ Bowick, P Mishra, R Sknepnek, M Cristina~Marchetti, Defect dynamics
  in active nematics.
\newblock {\em\protect\JournalTitle{Philosophical Transactions of the Royal
  Society A: Mathematical, Physical and Engineering Sciences}} \textbf{372},
  20130365 (2014).

\bibitem{tsai2008solutions}
CC Tsai, Solutions of slow brinkman flows using the method of fundamental
  solutions.
\newblock {\em\protect\JournalTitle{International journal for numerical methods
  in fluids}} \textbf{56}, 927--940 (2008).

\bibitem{ronning2022flow}
J R{\o}nning, CM Marchetti, MJ Bowick, L Angheluta, Flow around topological
  defects in active nematic films.
\newblock {\em\protect\JournalTitle{Proceedings of the Royal Society A}}
  \textbf{478}, 20210879 (2022).

\bibitem{hemingway2016correlation}
EJ Hemingway, P Mishra, MC Marchetti, SM Fielding, Correlation lengths in
  hydrodynamic models of active nematics.
\newblock {\em\protect\JournalTitle{Soft Matter}} \textbf{12}, 7943--7952
  (2016).

\bibitem{ambegaokar1980dynamics}
V Ambegaokar, B Halperin, DR Nelson, ED Siggia, Dynamics of superfluid films.
\newblock {\em\protect\JournalTitle{Physical Review B}} \textbf{21}, 1806
  (1980).

\end{thebibliography}

\end{document}